# Reliability Aware Multiple Path Installation in Software Defined Networking

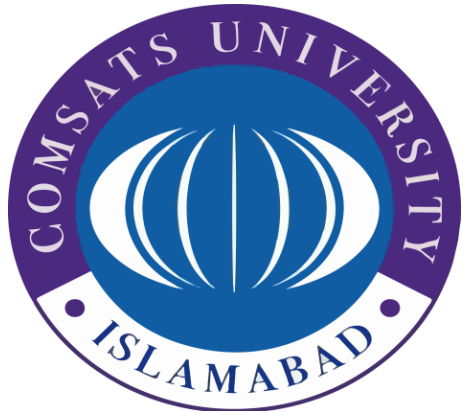

*By*

Syed Mohsan Raza

CIIT/FA17-RCS-006/Wah

MS Thesis

In

Master in Computer Science

## COMSATS University Islamabad, Wah Campus

Spring, 2019

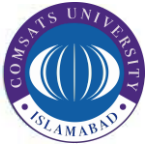

**COMSATS University Islamabad, Wah Campus**

# Reliability Aware Multiple Path Installation in Software Defined Networking

A Thesis Presented to

COMSATS University Islamabad, Wah Campus

In partial fulfillment

of the requirement for the degree of

## MS (CS)

By

Syed Mohsan Raza

CIIT/FA17-RCS-006/Wah

Spring, 2019

# Reliability Aware Multiple Path Installation in Software Defined Networking

A Post Graduate Thesis submitted to the Department of Computer Science as partial fulfillment of the requirement for the award of Degree of MS in Computer Science.

| Name | Registration Number |
|---|---|
| Syed Mohsan Raza | CIIT/FA17-RCS-006/Wah |

**Supervisor**

Dr. Nadir Shah

Associate Professor

Department of Computer Science

COMSATS University Islamabad,

Wah Campus.

July , 2019

# Final Approval

This thesis titled

# Reliability Aware Multiple Path Installation in Software Defined Networking

By

*Syed Mohsan Raza*

*CIIT/FA17-RCS-006/Wah*

Has been approved

For the COMSATS University Islamabad, Wah Campus

External Examiner: _______________________________

Dr. _________________________

Supervisor: _Dr._Nadir_Shah_________________

Department of Computer Science, CUI, Wah Campus

HoD: __Dr._Muhammad_Wasif _Nisar _____

Department of Computer Science, CUI, Wah Campus

# Declaration

I **Syed Mohsan Raza** hereby declare that I have produced the work presented in this thesis, during the scheduled period of study. I also declare that I have not taken any material from any source except referred to wherever due that amount of plagiarism is within acceptable range. If a violation of HEC rules on research has occurred in this thesis, I shall be liable to punishable action under the plagiarism rules of the HEC.

Date: \_\_\_\_\_\_\_\_\_\_\_\_\_\_\_\_\_ Signature of the Student

\_\_\_\_\_\_\_\_\_\_\_\_\_\_\_\_\_\_\_\_\_\_\_\_

Syed Mohsan Raza

CIIT/FA1-RCS-006/Wah

# Certificate

It is certified that Syed Mohsan Raza having registration number CIIT/FA17-RCS-006/Wah, has carried out all the work related to this thesis under my supervision at the Department of Computer Science, COMSATS University Islamabad, Wah Campus. The work fulfills the requirement for award of MS degree.

Date: _________________

                                        Supervisor:

                                        _______________________________

                                        Dr. Nadir Shah
                                        Associate Professor
                                        Department of Computer Science, CUI, Wah Campus

Head of Department:

_______________________________

Dr. Muhammad Wasif Nisar
Associate Professor
Department of Computer Science, CUI, Wah Campus

To Almighty ALLAH and the Holy Prophet Muhammad (P.B.U.H)

&

My Loving Parents, Teachers, Brother and Friends

# ACKNOWLEDGEMENT

In the name of "ALLAH" The Most Gracious, The Most Merciful. To him we belong, and all the praise and thanks are due. After thanking Almighty "ALLAH" for his blessing and guidance to complete this work, I would like to express my sincere gratitude to my supervisor, **Dr. Nadir Shah**, for providing me valuable advices and guidance in every stage of my degree program. Throughout the training under his guidance, I established solid practical and problem-solving skills, which prepared me for any challenge in my future career.

I also want to render gratitude to my teachers and team members, Mr. Mudassar Hussain, and Mr. Rashid Amin for their understanding, patience, and support throughout my studies. I can never forget the beautiful moments of happiness in their relaxing company during a hard time of this degree. I can never thank enough for everything they did for me. At my end am thankful to all of my teachers, real mentors of my past, from first day of learning to till day. My parents support for study is unique and there is nothing to return in this regard. Thanks to all other team members working in collaboration and they are members of SDN research group under supervision of respected supervisor.

**Syed Mohsan Raza,**

**CIIT/FA17-RCS-006/Wah**

# ABSTRACT

# Reliability Aware Multiple Path Installation in Software Defined Networking

Being a state-of-the-art network, Software Defined Networking (SDN) decouples control and management planes from data plane of the forwarding devices by implementing both the control and management planes at logically centralized entity, called controller. This helps to make simple and easy both the network control and management. Failure of links occurs frequently in a computer network. To deal with the link failures, the existing approaches computes and installs multiple paths for a flow at the switches in SDN without considering the reliability value of the primary path. This incurs extra computation to compute multiple paths, and both increased computation time and traffic to install extra flow rules in the network. In this research work we propose a new approach that calculates the link reliability and then installs the number of multiple paths based on the reliability value of the primary path. More specifically, if a primary path has higher reliability then a smaller number of alternative paths should be installed. This shall decrease the path computational time and flow rule installation load at controller. Resultantly there shall be less flow rule entries in switch flow table which in turn will avoid the overflow of the flow table. Through simulation results, our proposed approach performs better as compared to the existing approach in term of computational overhead at controller, end-to-end delay for packet deliver and the traffic overhead for flow rule installation.

# Table of Contents

# List of Figures

# List OF ABBREVATIONS

| | |
|---|---|
| ACL | Access Control List |
| AP | Alternate Path |
| FF | Fast Failover |
| FLI | Fault Location Identification |
| ICT | In-band Control Tree |
| PAM | Path Aliveness Monitoring |
| OVS | Open Virtual Switch |
| RAF | Reliability Aware Flow Installation |
| SDN | Software Defined Networking |
| TCAM | Ternary Content Addressable Memory |

# Chapter 1
# Introduction

# 1 Introduction

In traditional network, both the control and management planes are implemented along with data plane at each network device (e.g. switches/routers). This causes many problems like network oscillation problem, misconfiguration due to manual configuration of each network device using low level commands and closed operating systems of forwarding devices, etc., [1][2]. To address these problems, Software defined networking (SDN) [3] is emerged as a new architecture that physically separates the control and management planes from the data plane of forwarding devices by implementing both the control and management planes at a logically centralized entity, called SDN controller [4]. This makes easy to both control and manage the network. Using SDN controller, a network operator can efficiently implement network configurations like load balancing, traffic engineering security enforcement, etc. Central control in SDN makes the network maintenance operation simple which include the reachability map, enforcing access control list (ACL) at a single controller. The SDN architecture consists of three planes, as shown in Figure 1.1. Application plane, control plane and data plane. Application plane communicates with control plane using Northbound API. Data plane is controlled by the decision made at control plane using southbound API (most often the OpenFlow [5,6] protocol is used). Networking devices (like routers, switches, etc.) and hosts explicit register themselves with controller and periodically update the controller about its link state information. Thus, the SDN controller has global view of the network. This makes the SDN more transparently manageable. [7.]

## 1.1 Application Plane

Application or management plane is set of end user applications that interact with the control plane, like security applications are responsible to counter attacks and threats, load balancer is used to evenly distribute the traffic among the links, SDN programming languages (like Frenetic[8]) are used to specify the configuration and requirements at abstract level, etc.. Application plane is the interface for the administrators to develop applications and customize behavior of network. This entity makes network programmable and flexible for developers.

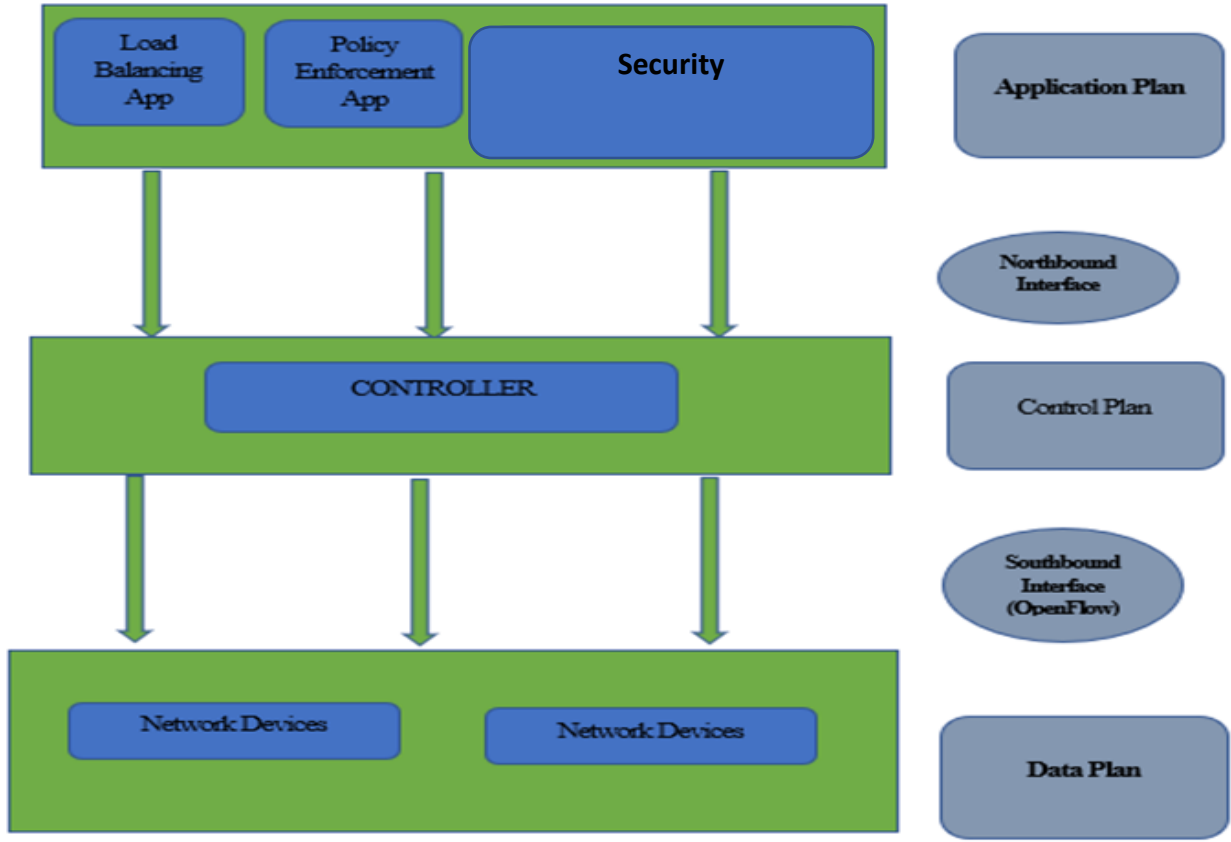

*Figure 1.1-SDN Architecture*

## 1.2 Control Plane

SDN control plane is the decision logic which specifies the action for a data packet based on application plane and network topology. The action for a data packet can be to forward through a specific path, or drop or send to controller. The controller communicates the decision to the data plane by installing the flow rules at the switches along the path. Controller has awareness of network topology. Control plane can be run over a single machine or multiple machines.

## 1.3 Data Plane

Data Plane in SDN consists of a set of forwarding devices (Open switch, routers, Access points) that are commanded by the controller using OpenFlow protocol to take action for a data packet arrived at the switch. A data plane establishes a secure connection with SDN controller and has a flow table which consists of pattern and action fields. When a data packet is arrived at the switch, the switcher looks for the matching pattern in the flow table., If the matching is found, then the switch takes the corresponding action. Otherwise, the switch shall communicate with controller via OpenFlow to compute the action for the data packet.

## 1.4 SDN Advantages

Software Defined Networking has several advantages competitively to traditional IP network. It makes easy to manage the network. Administrator can change the network policies easily form a single location (i.e. controller) and distribute it to all forwarding device while needed. In traditional network, it is hard to localize the failure in large network. However, in case of SDN, controller has dynamic and instant awareness of failure location. SDN paradigm provides open programming application interfaces for easy application development that interact with SDN planes. [9Due to the numerous advantages, SDN has been adopted by many organizations, e.g., ONF [10], AT&T, NEC, and Huawei Technologies [11]. Google [12], Microsoft and VMware [13] [14]. Most of organizations are adopting the SDN along with their already functional network structures.

## 1.5 SDN Challenges

Beside these benefits, SDN has also several challenges as follows. (i)Controller placement: every switch is continuously communicating with controller. So there is a need to place the controller at optimum place to minimize the load at controller, delay toward the controller, and the distance to controller from each switch. [15].
Similarly, the communication between application plane and control plane, and control plane and data plane should be standardized and open source.
 Link Failure resiliency is also a challenge to provide better quality of service to the users despite the link failure. Link failure occurs frequently in any computer network. This degrades the network performance if a link remains failed for longer time without taking measures for its recovery.

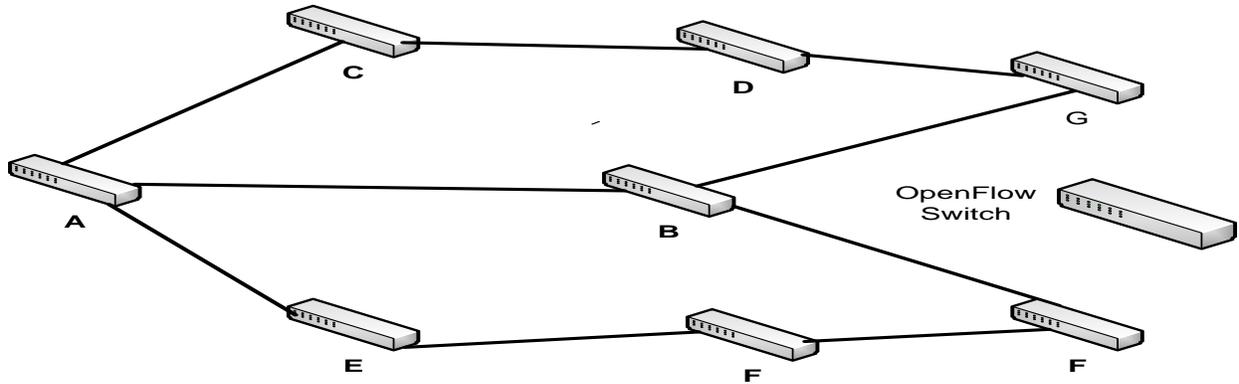

*Figure1(a) -Supposed Network View for Proactive and Reactive flow Rule Installation*

The existing approaches for link failure in SDN can be categorized into two types [16,17]; proactive and reactive approaches. To explain these approaches, suppose we have a network as shown in Figure 1(a).

(i)-**Reactive Failure Recovery Mechanism [Restoration]**

When a link gets failed at the switch, the switch informs the controller by sending the link failure event and asks the controller to compute another alternative path for the flows passing through the failed link. After this, controller computes and installs the alternative paths for all the flows passing through the disconnected link. Then the data packets these flows start to be forwarded along the new paths. The disadvantage of this approach is that it introduces longer delay taken during the process to inform the controller and subsequently to compute and to install the alternative path. For example, there are two paths between switches A and G in the figure. If the link between switches A and B is failed for the flow between switches A and G, then the switch A will inform the controller and the controller will compute another path (A-C-D-G) as shown in Figure 1(b).

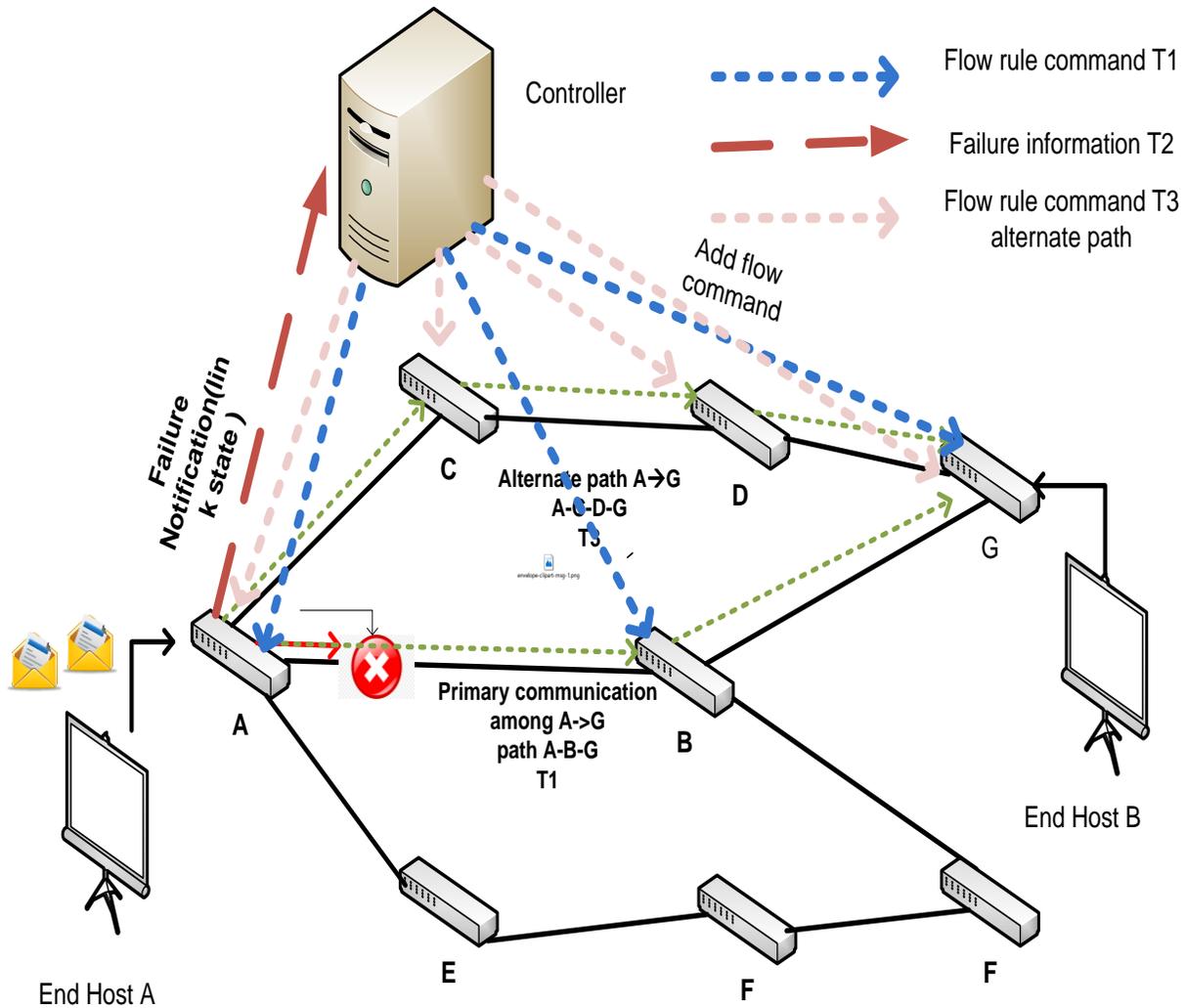

*Figure 1(b)- Reactive Alternate path installation in case of failure*

**(ii)-Proactive Failure Recovery Mechanism [Protection].**

The controller installs multiple paths at a switch for each flow. When a switch detects a link failure, the switch routes the flow over the alternative path without contacting controller. For example, in Figure 1(c), the controller computes two paths for a flow passing between the switches A and G and installs two paths for example (A-B-G) and (A-C-D-G) proactively.

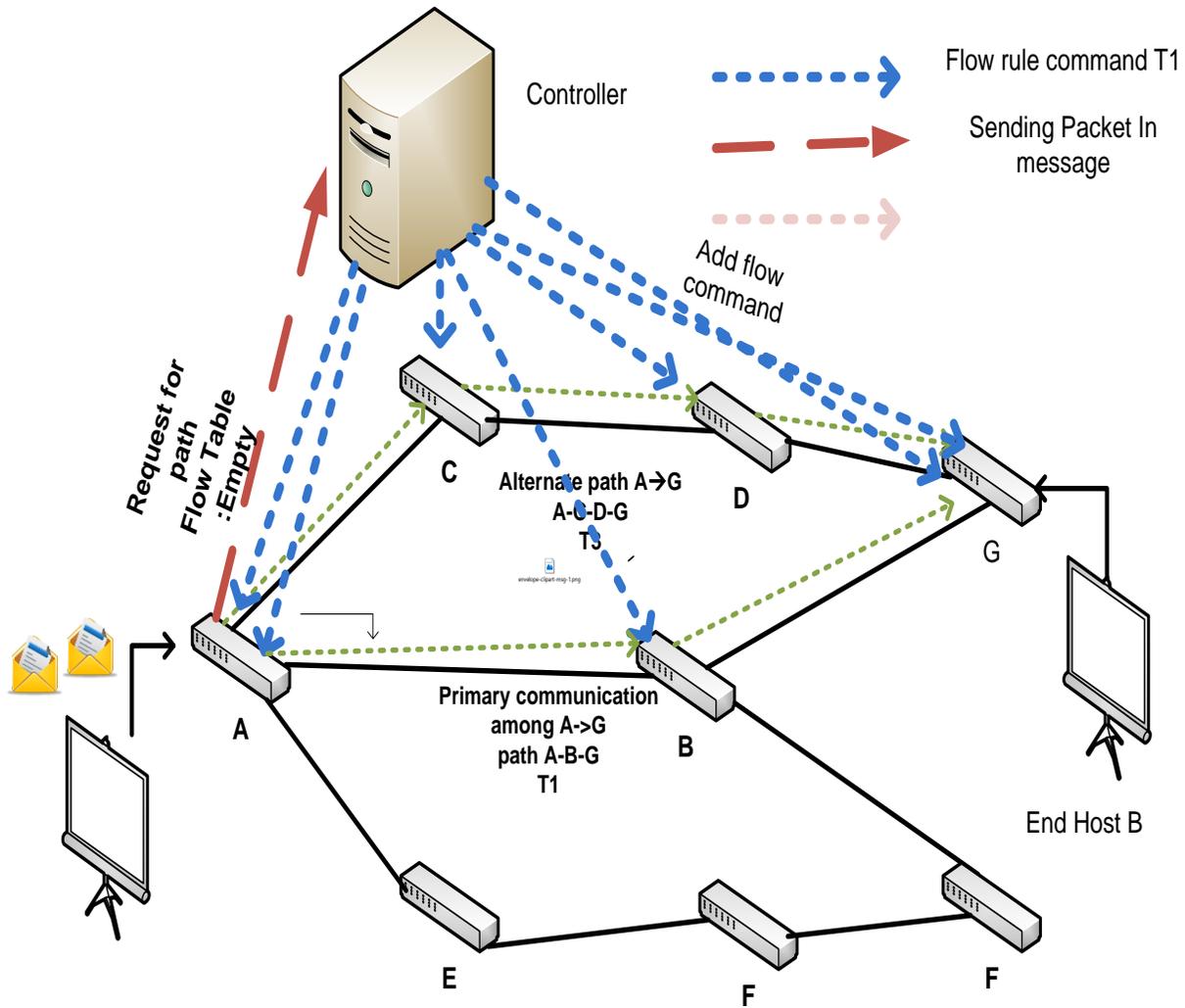

*Figure 1(c)- Two Path Installation by Controller Proactively*

Thus, this mechanism is faster than the reactive failure recovery mechanism because of less controller intervention involved in this process [18-20]. This mechanism has associated disadvantages of longer delay to computing multiple paths, larger traffic overhead to install multiple paths at the switches, and consumes more memory space at the switches for storing multiple paths in the forwarding table. Moreover, the SDN switches have very limited Ternary Content Addressable Memory (TCAM). Approximately, 1500 flow rules could be stored in Ternary Content Addressable Memory (TCAM) [21,31] of a SDN switch [22]. TCAM is an expensive (in term of its cost) and faster memory. The existing approaches for SDN stores multiple paths for every flow at each switch regardless of the path reliability. This can overflow the TCAM memory of the switches. More specifically, if a path for a flow is reliable (i.e. its failure chance is very low) then there is no need to install multiple paths. This will reduce the computation time at the controller for computing single path, traffic overhead by installing the flow rules (path) along the single path, and the TCAM memory consumption by having single flow entry in the forwarding table of the switch.

Our proposed approach, called RAF, solves this problem formerly as follows. In our proposed approach each switch periodically exchange the link failure information along with other link state information with the controller. Then the controller computes the reliability of each link. After receiving a request for path computation for the flow, the controller computes the number of multiple paths based on the reliability level of the primary path. More specifically, if the primary path for a flow is most reliable (in our proposed approach if the reliability of the path is greater than 90%), then the controller computes and installs single path. If the reliability level of the primary path is between 80% and 90%, then the controller computes and installs two paths for the flow at the data plane. More detail of the proposed solution is given in Chapter 4. Through simulation results, we show that our proposed approach performs better than the existing approaches in term of end-to-end delay, traffic overhead and computation time overhead.

Rest of the thesis is organized as follows. A comprehensive overview of related literature work is described in Chapter 2. Chapter 3 explains the problem statement through an example scenario. The detail of our proposed approach is given in Chapter 4. Chapter 4 describes the simulation results. Finally, Chapter 5 concludes the thesis along with some future research directions.

# Chapter 2

# Literature Review

Many researchers have provided different solutions for link failure handling, the detail is as followed.

The authors in [18] States that Fast Failover group (FF) is an OpenFlow switch specification used for detouring the flows to alternate port of OpenFlow switch, when a link failure occurs and switch request to controller. In this mechanism, flow rules have group ID in their action filed. This group ID invokes the group table entries. Group Table entries forward packets to switch port defined in action bucket. When failure occur next available action, bucket is activated, and status of previous action bucket is disabled. Figure 2 explains how flow rules uses primary and alternate action bucket in group table, when port 2 status is down the successor action bucket shall divert packet to port 4 of switch for immediate recovery. The authors suggest a new technique, called Controller Independent Proactive (CIP) that uses flow grouping technique and forwarding rules aggregation technique as follows. In CIP flows at a switch that has same output ports are placed in one group. In forwarding rules aggregation, the VLAN [23] tagging is performed to the disrupted flows to reduce the memory storage consumption in the switch of alternate path.

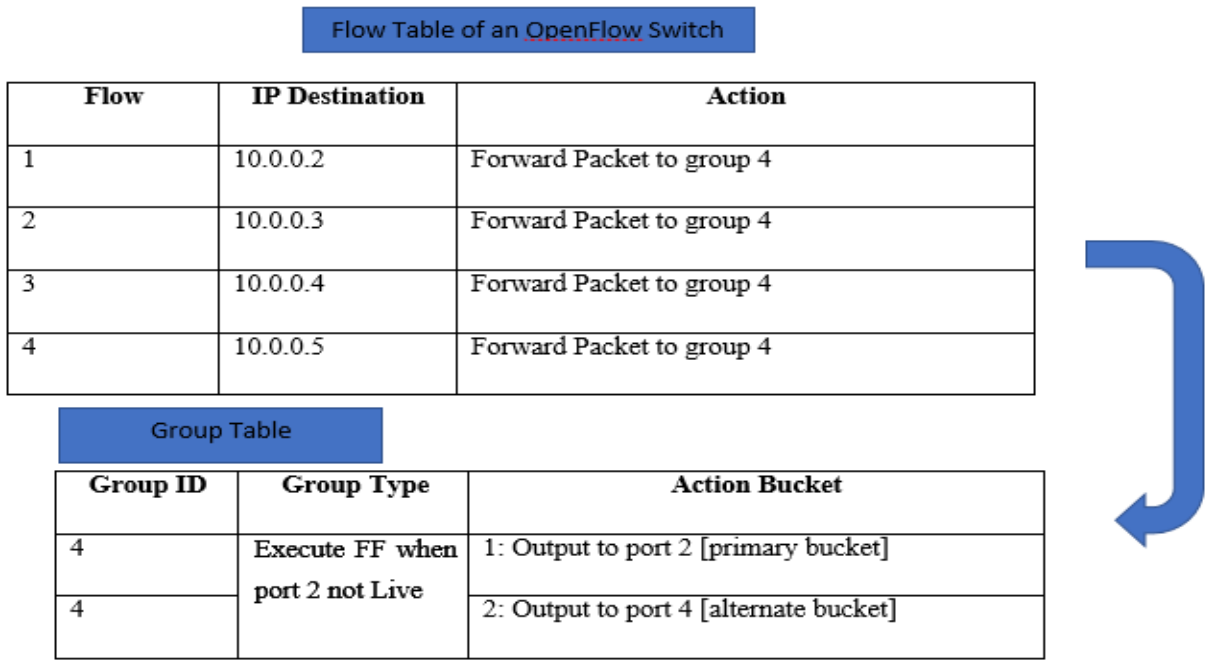

*Figure 2.1-Group Table entries and Action Bucket*

Lee et. al. [24] proposes a mechanism of fast failure detection which is applicable both for in-band and Out-of-band OpenFlow networks. Out-of-band networks are widely used because these are easy to manage but Controller need a dedicated port of each switch for control channel. While in-band Control does not need port reservation. In-band network failure resilience in hard to manage because when a single link failure occur it cause many data and control channels disruption. Figure 3 presents example of in-band and out-of-band networks

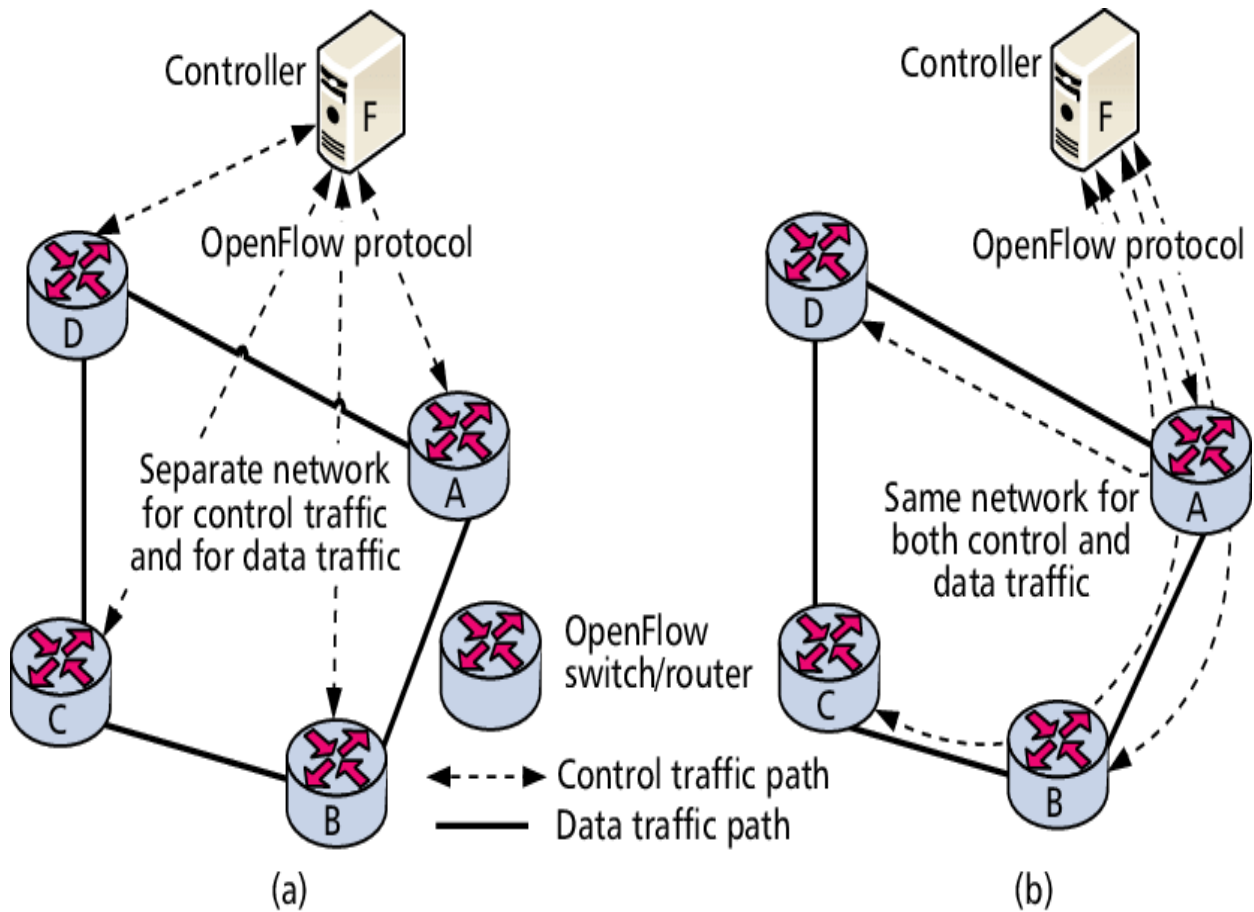

*Figure 2.2-In-band and Out-Of-Band OpenFlow Networks*

Author identifies problem of time taken for link failure detection involves recovery delay. Failure recovery depends on link failure detection and rerouting the disrupted flows. Although proactive recovery techniques provide immediate recovery but if long time is required by OpenFlow switch for detection of a link failure, recovery time and number of dropped packets in network shall also increase dependently. In order to address this problem, the authors suggest to use monitoring cycle technique, a software-based failure detection technique that detects and localizes the failure in short time. Monitoring cycles include controller and switches in cycle. Time of recovery using monitoring cycle is comparatively evaluated with Bidirectional Forwarding Detection [25] technique in this paper. Monitoring cycles has two rounds.

(i) Path Alive Monitoring (PAM) in which controller forwards a monitoring packet in cycle. If all link in path are alive, Controller receive ok message.

(ii) Fault Location Identification (FLI) round starts as a link failure occurs, and consequently controller does not receive the monitoring packet in a predefined time FLI round pinpoint the link failure. In in-band network these two rounds are not enough for link recovery. In-band Control Tree (ICT) is maintained of network devices in network normal operations.

A single link failure can disrupt multiple control and data channels. PAM and FLI compute either data or control channel disruption in-band network. If control channel fails, then ICT is divided. Controller modifies the ICT using proactively computed multipath and reestablishing controller to switch reachability. Results of this work show that fifty milliseconds time of monitoring the link can yield the recovery time less than one hundred milliseconds in an in-band OpenFlow network in its worst case. Figure 3 illustrate working of both PAM and FLI rounds in out-of-band network.

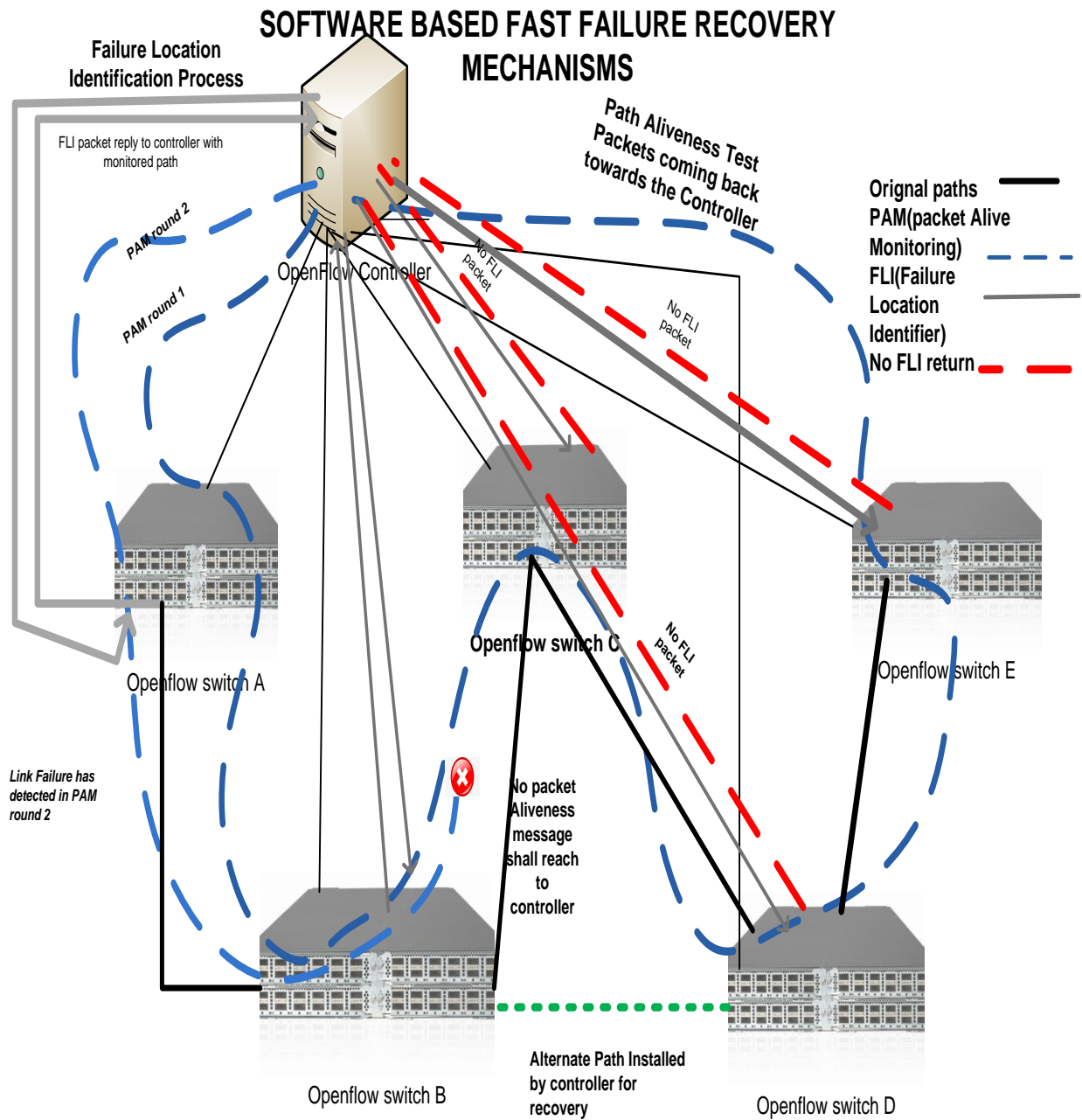

*Figure 2.3- Path Aliveness Monitoring and Failure Location Identification in Software-Based Fast Failure Recovery*

Automatic Failure Recovery [AFRO] is proposed in [26] that conceptually decouples the controller operations from failure recovery mechanism. AFRO address a problem of application complexities. Application running in control plan need to include failure logics. There are some Controllers POX, NOX and Floodlight [27] e.g., based on modular programming, this make more difficult to include recovery logics in each software module. Author explored that recovery simple logic to mitigate failure in Application plan like time out and deleting the forwarding rules on the base of inactive port is insufficient. Moreover, such action could be cause of bugs, loops and harder to retain transparency of control logics. Controller contribution in very simple state forward technique could be as when a failure or topology change occur just to clean all states and restart the Controller. This can normalize the traffic flow. It has associated disadvantages as multiple packet drop and time taken to clear all entries form switch tables, resetting the controller and network, re computing path and installing the rules.

AFRO use the soft coded logics of controller, such that it records all the controller behavior in normal execute in its record phase. Mainly recording encompass the Flow mode installation and removal in actual network. In case of failure occurrence, a shadow controller is activated. This shadow controller has same functionality as original ones but it installation take place for a clean state of flow entries. When Shadow Controller is activated, Shadow network for this Controller is also formulated by removal of failure elements from the original network. AFRO replays the recorded actions of original controller using shadow controller. This mainly focus that how the PackectIn messages was processed before the failure occur. Upon completion of replay phase, Shadow Controller and network carry new forwarding rules. After this reconfiguration transition to original network take place in which there are two methods. First directly take over by shadow Controller to original network. Second one is by the modification of flow rules form the switch tables such that targeted rule removal and new computed rules installation. During this procedure traffic is directed from previous to new rules by putting a barrier for traffic to wait provided a switch is reconfigured. Figure 5 shows the working of Controller and Shadow controller in order to execute the recorded rules before failure and reconfiguring the switches when a link failure occur.

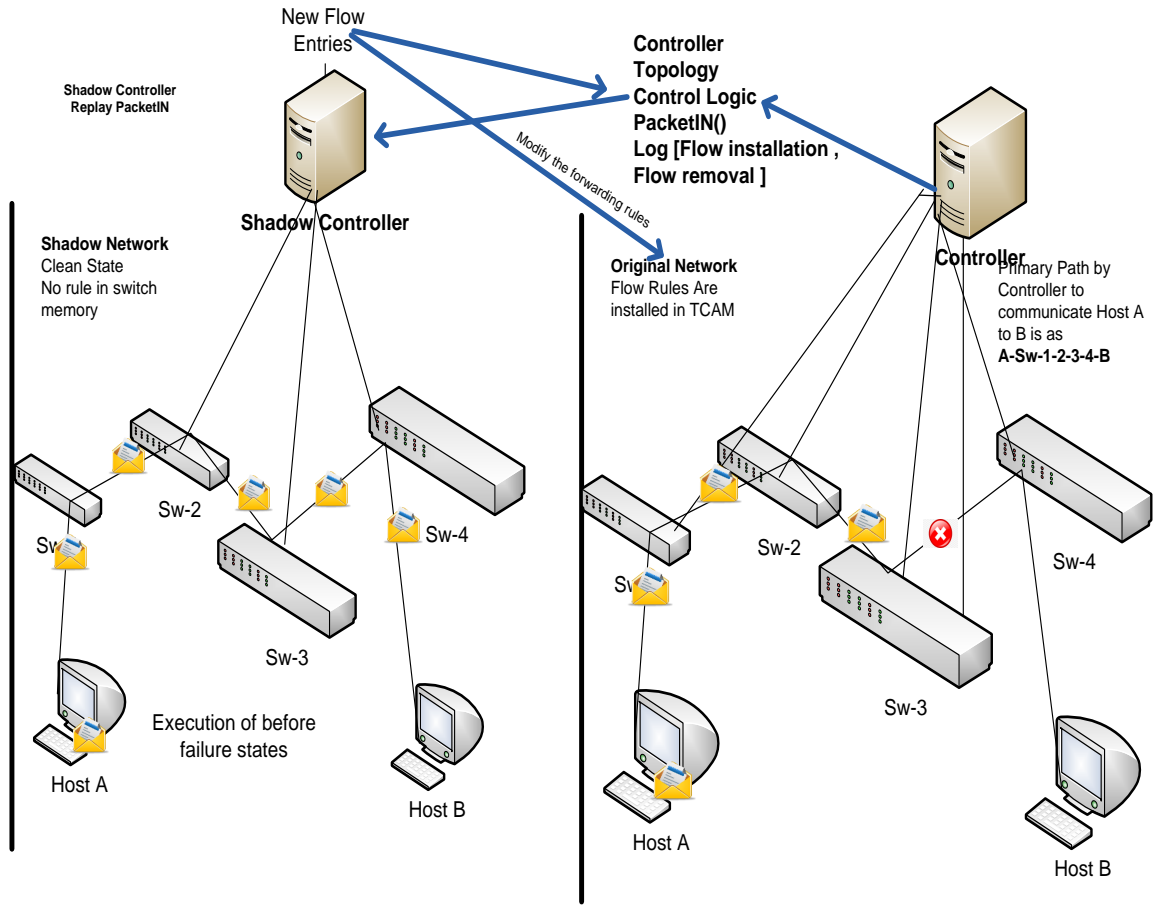

*Figure 2.4- AFRO recovery mechanism [Actual + Shadow Network]*

Lin et. al. purpose switchover mechanism in [28] to tackle the link failure problem in software defined networks. Author identifies a problem of Fast failover technique that when multiple flow passing through a switch has same output port are directed towards a live port using action bucket in group table, this port set congested in this case. Main contribution of this paper is switchover mechanism that address the port congestion problem by swapping the action buckets and configuring the less transmission status port as live Port dynamically. Swapped ports belong to backup paths installed by controller in switch tables proactively. For the purpose Controller store information of ports statists of every switch by periodic statistics requests. A threshold is pointed for transmission rate . if the average transmission rate of a switch egress port is continuously higher than the threshold then controller shift the flow from primary to backup path port. Controller get the switch interlinks via PackectIn message. Controller receive link information using LLDP And host to connected switch link information has extracted via northbound interface (Rest API) in simulation of this work. Flow switchover and congestion localization are two stages of fast failover strategy. When Controller measure the port congestion it set its status as congestion port otherwise it's a normal port . In fast switch over there should be parallel support of Fast Failover because when Controller find a transmission rate elevated from threshold it checks the Action Bucket pointed by Group ID. In flow table these group ID are action to flow a rule. When controller find an alternate port in further bucket with normal status it swaps flow toward that egress port. Simulation results of this paper argue that it has efficient failure recovery period than that of straight forward protection mechanism.

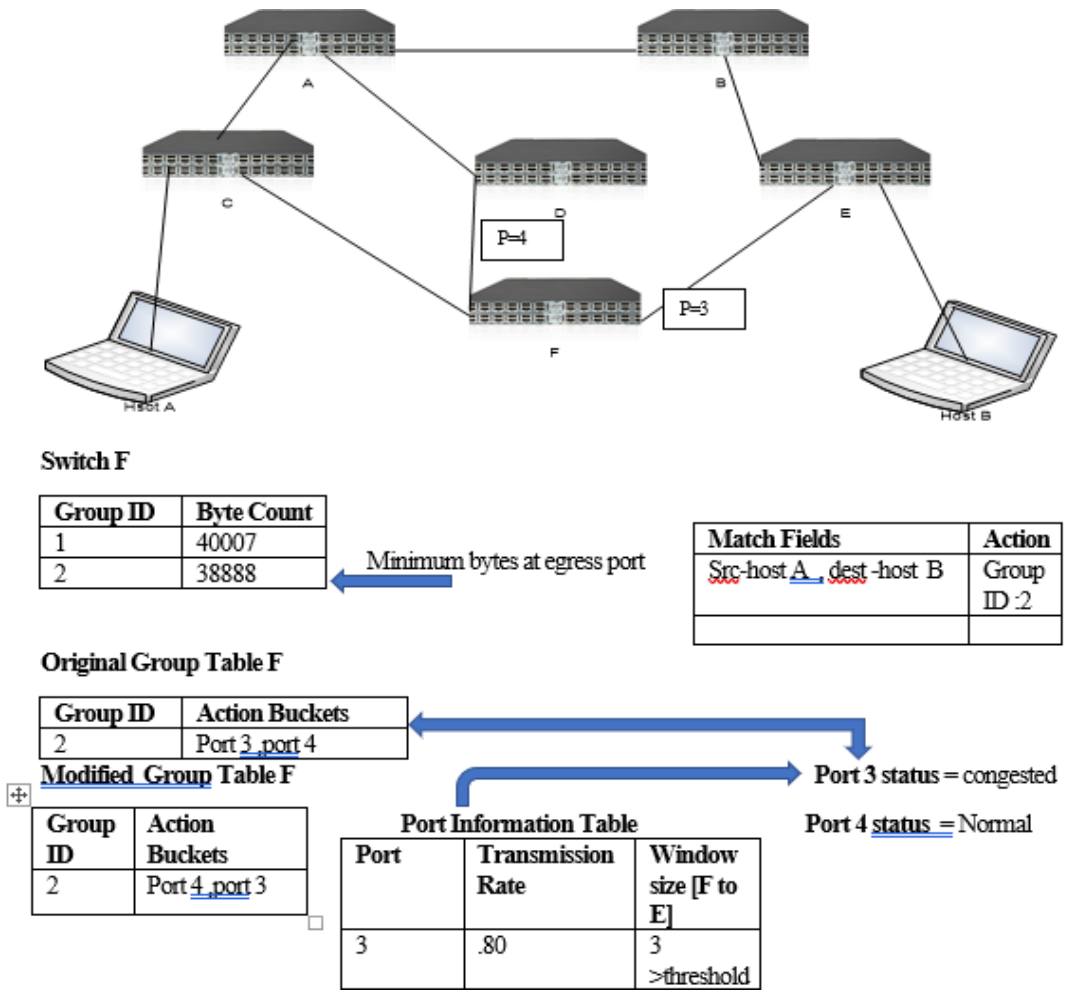

*Figure 2.5- Flow Switchover mechanism to activate the backup path by Action Bucket ports modifications*

Figure 6 explain the working of switchover in which switch F has flow table and a group table. First live port in action bucket of group entry is 3. Just after accessing the port status congested controller modify the group table using OpenFlow group modification command and normalize the flow.

CORONET [29] is a system of Fault tolerance of data plan in SDN Authors of this paper sates problem such that recovery technique of failed path using the straight forward Fast Failover, flow aggregation for detouring the disrupted flows are advantageous but in case if SDN architecture has deployed in a large-scale network, there are thousands of flows per link. Such case bottlenecks the centralized controller secondly such velocity of flow rule installation in switch memory decrease the switch resources utilization. More specific, if controller performs extra calculation for flow rule modification then link failure recovery time is elevated. failure flow modification in gradually increasing network is cause of increasing the recovery time. This system provides the solutions of recovery and address problem of multiple link failure in data plan in SDN. CORONET has a central decision point. Because of local switch mechanism for link testing using LLD packets, system has less control interception in recovery mechanism. Route planning module is responsible to compute multiple routes on base of updated network topology information. In this work SDN application can use only the logical path configured by OpenFlow Protocol during its VLAN switch configuration module. In this module multiple ports are configured with VLAN IDs. After this host traffic is forwarded to compute logical path randomly or by using round robin algorithm. Modules involved in CORONET Fault tolerance system are elaborated in Figure 7.

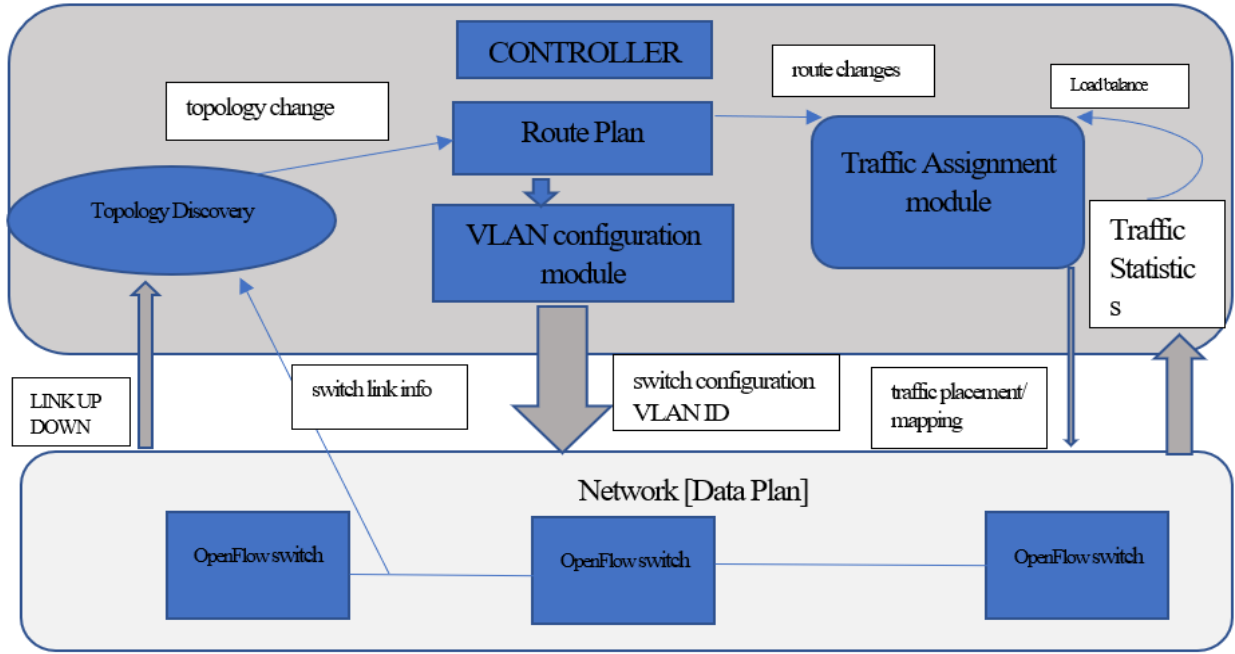

*Figure 2.6-Coronet Controller Architecture*

CORONET features enable to work this system with any type of topology. Controller porotype is developed on the top of NOX Operating System. NOX runs multiple control process and knows network devices linkage. Moreover, this single view of network is aware about the location of host, host address and all of bindings. When a packet has arrived at switch form some of source towards a destination and there is no flow entry available for pair of host then its referred to NOX process to get path. Mostly first packet is received at controller. An event driven approach has been used in NOX development because Enterprise network topologies are not static. Addition to its explanation is as host of networks are added and removed. There are some events generated by OpenFlow itself. Switch join, leave , switch feature and stats request are examples [30].

Yu et. al. [31] Proposed a link failure framework. This framework provide the resilience against the link failure using OpenFlow in existing routers. Routing process generate the routing information database while OpenFlow controller configure flow table for communication. So, there are two layers of forwarding structure. OpenFlow enabled forwarder are to forward data by searching in flow table and taking the corresponding action of flow entry .If relative rule does not exist for source to destination in flow table then routing information database in followed. Detection program is a component of forwarding node in case of link failure occur. When link failure occur, controller is informed about failure state and instructions to compute the backup path. During period of link failure, the routing database and flow table work as collaborator. Packets affected by link failure shall be process using the flow table. For the purpose , Controller find the alternative paths using the Floyd algorithm ignoring the failed links. But it's a post failure recovery or reactive action performer for recovery purpose not a proactive approach like Veriflow , a layer between controller and data plan to evaluate the flow rule proactively before installation [32].

SPIDER [33] is a pipeline packet processing that apply the stateful design for the recovery mechanism in switch locally. It provide the functionality using fully customizable and programmed techniques for recovery directly in switch. As compare to other technique SPIDER does not rely at Controller intervention in Data plan of SDN network for recovery purpose. It is mainly composed of Open State. In which state tables are used flow table are preceded by the state tables. When a packet is for processing using flow table entries, it is first matched to state available in table using a key generation by flow table. In case if there is no matched state ,table return the default state. Flow table update the state by set state method using unique key. Failure detection is by using heartbeat bidirectional packets among nodes of topology. If reply of this request does not return any packet to failure , port status is set to down. When a port is declared as a down port than any other port will be active to forward data. Using this mechanism SPIDER, guarantee the failure recovery in very short time. Behavioral Model in this work is in form of Finite State Machine which is acted for per flow processing.

Flow Table Compression in proposed in [34] to Utilize the OpenFlow switch memory efficiently. This solution addresses the problems of Data center networks as in datacenters link reliability, multipath availability and immediate recovery is requires. To insure these requirements when a switch has multiple paths along with a primary path in its memory, this mechanism cause to increase the inefficient usage of TCAM usage that is a crucial resource. Compression is performed for flow entries that have same action outputs in a switch locally. In this way a Flow table get a smart entries. This work also provides the forwarding on base of compression awareness.

In [35] Dcell is network structure desired in data center network. It provides the fault tolerance not for a single link failure. Recovery protocol is distributed in this structure. When there are many failure in data center network Dcell calculate the communication path. This path has very low distance from primary shortest path . High level Dcells are the combination of multiple low level Dcell. It run fault tolerant protocol [DFR] at the top of interconnection. Ultimate destination of Dcell is scaled data centers. Fast Failover group is an Open Flow blessing to test the port liveness. Group ID contain a live port in its each associated action bucket. Buckets are executed as defined already in switch. In case of port status down the successor bucket is activated [36].

From the above presented literature and rest of considered failure resiliency approaches in SDN [36-44] one can concludes that none of them installs the number of multiple paths based on the reliability level of the primary path for a flow in SDN.

# Chapter 3
# Problem Statement

A number of techniques have been discussed in related work for proactive path installation to cope with the link failure. These studies reveal that alternate paths computation produces overburden at controller in large networks and inefficient use of switch resource (memory) at switches. The researchers have proposed a variety of mechanisms for the memory utilization of OpenFlow switches and reduction of controller involvement for failure recovery.

Beside this none of research work still has a mechanism to compute the reliability level of primary path at controller, and then based on reliability value of the primary path the number of other alternate paths for the primary path should be computed and installed in the network. For example, if a path is 100% reliable than the alternative paths should not be computed and installed for the path. Otherwise, the number of alternative paths for the primary paths should be based on the reliability value of the primary path. To explain the problem through an example scenario, suppose we have a network with given reliability level for each link in Figure 3.1 (a).

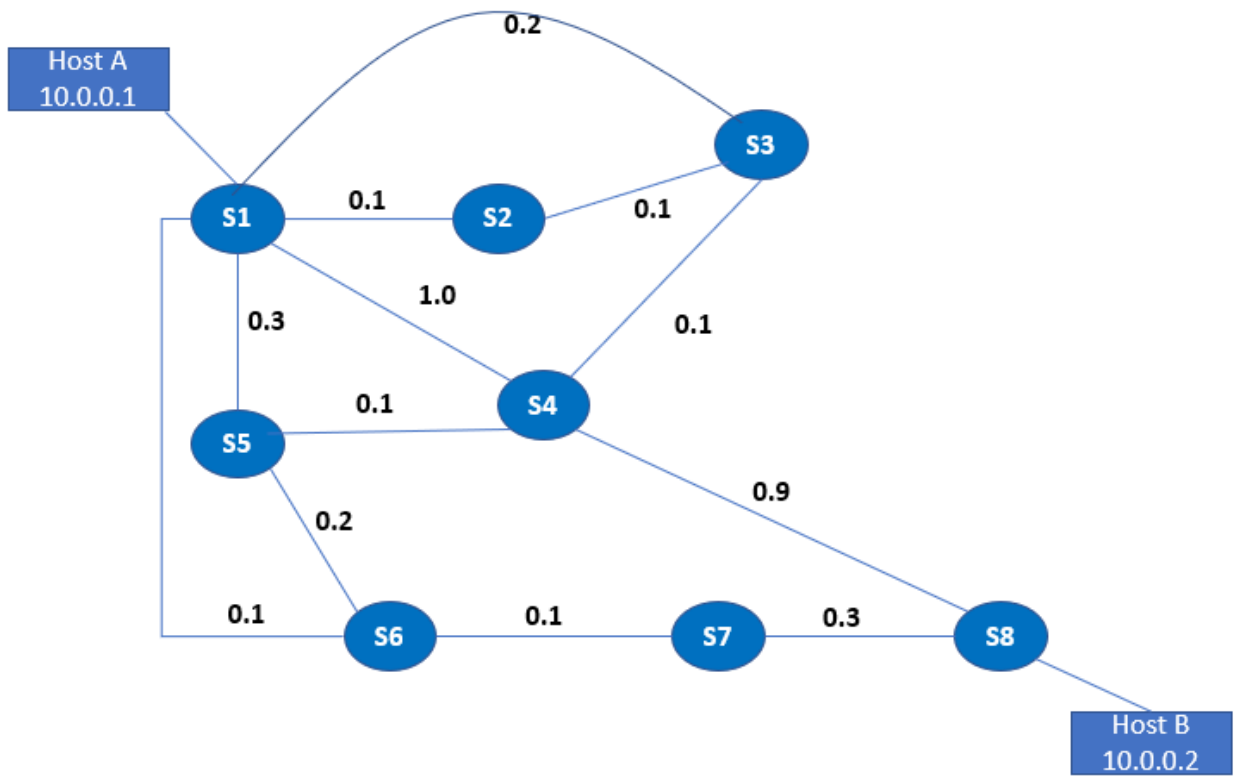

*Figure 3.1(a) Switches in topology and reliability weightage within corresponding edge*

The number of alternative paths between s1 to s8 are 10. If host A sends a packet to host B, then the packet will be received at switch s1. The switch (s1) shall send a packetIn toward controller, suppose the flow rule is not installed in the forwarding table of s1. Controller computers the primpary path (s1-s4-s8) which has maximum reliablity (0.9 i.e 90%) . In this case, the existing approach [44] will compute all alternative paths and will install all the paths in the network, as shown in Figure 3.1 (b). As the reliability is high (90%) of the primary path, so the other alternative paths installed are redundant. We suggest that in this case only the primary path should be installed, as shown in Figure 3.1 (c). This will reduce the compuation overhead for computing multiple paths at the controller, traffic overhead to install the flow rules at all alternative paths in the network, and memory usage in the switches.

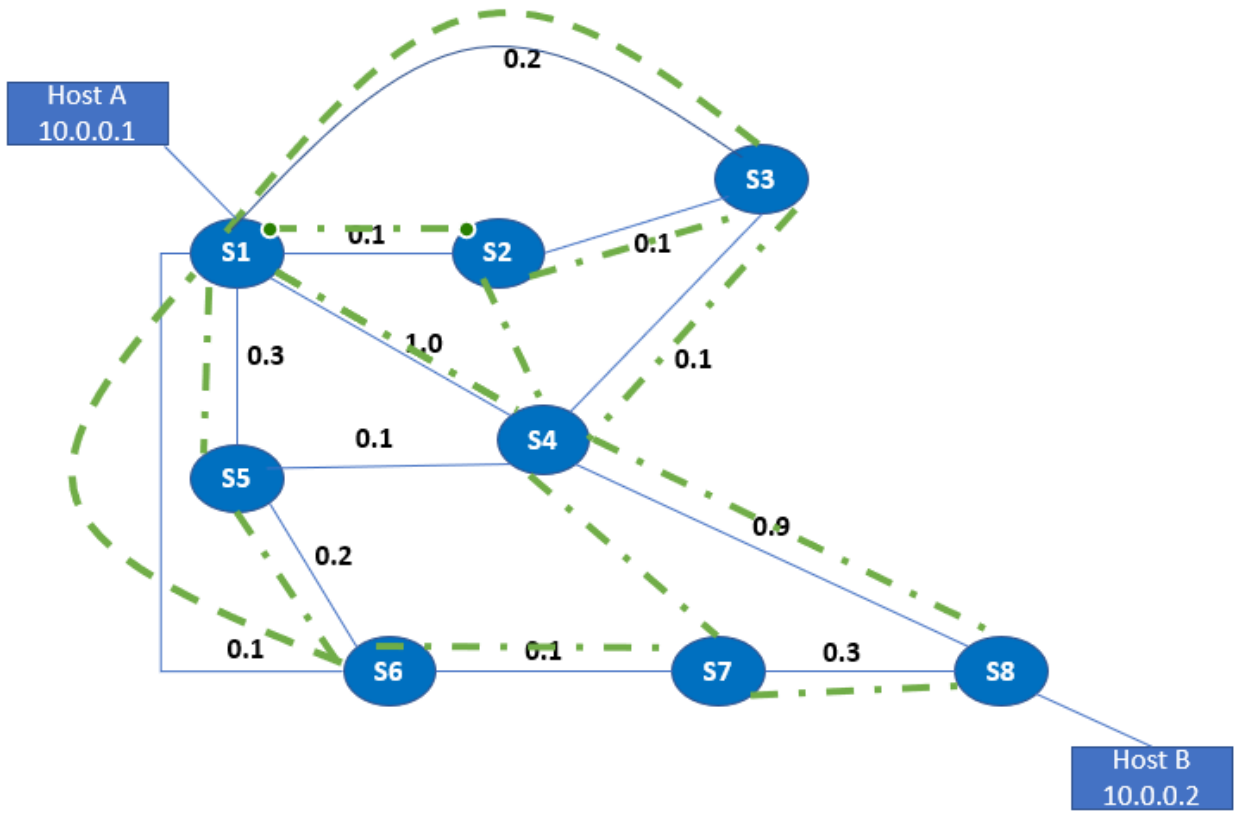

*Figure 3.1(b)-Flow Rule installation for all path*

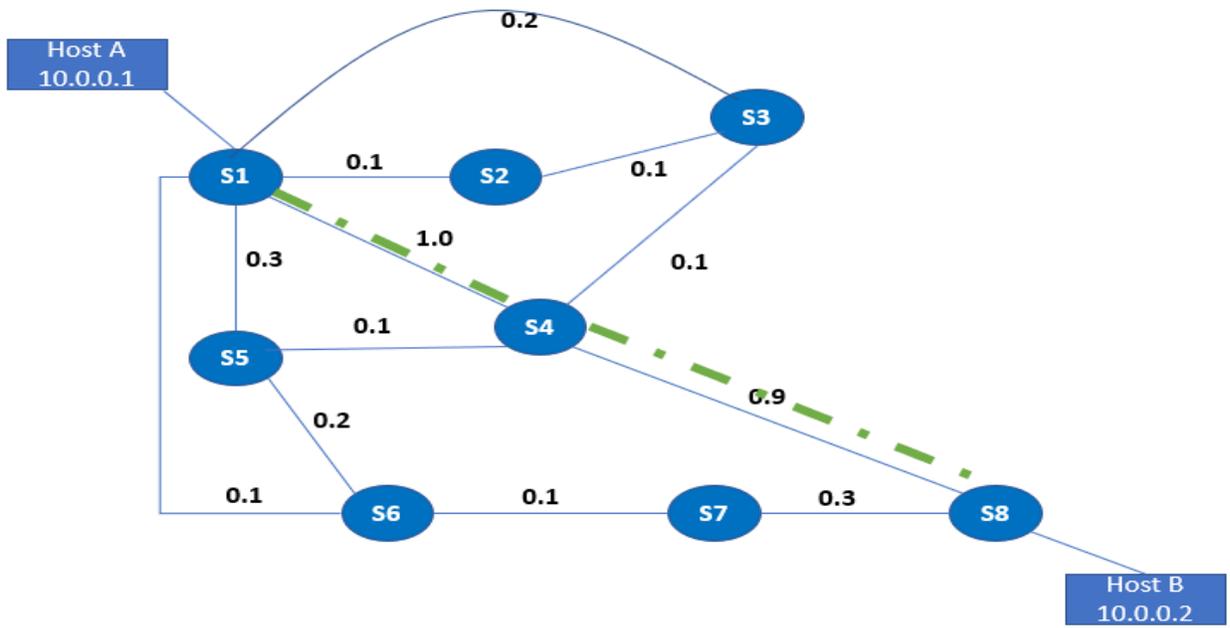

*Figure 3.1(c)-Single Reliable Path Installation*

# Chapter 4
# Proposed Methodology

## 4.1 Proposed Approach

In this section, we formulate our proposed approach, called Reliability-Aware Flow installation mechanism (RAF), that the primary path based on higher reliability. Then, we propose a variation of the RAF, Distance based RAF, that computes the primary path based on the joint value of higher reliability and shorter distance.. We assume out-of-band communication model for communication between a switch and controller, and reactive flow installation mode. Our proposed approach contains the following components.

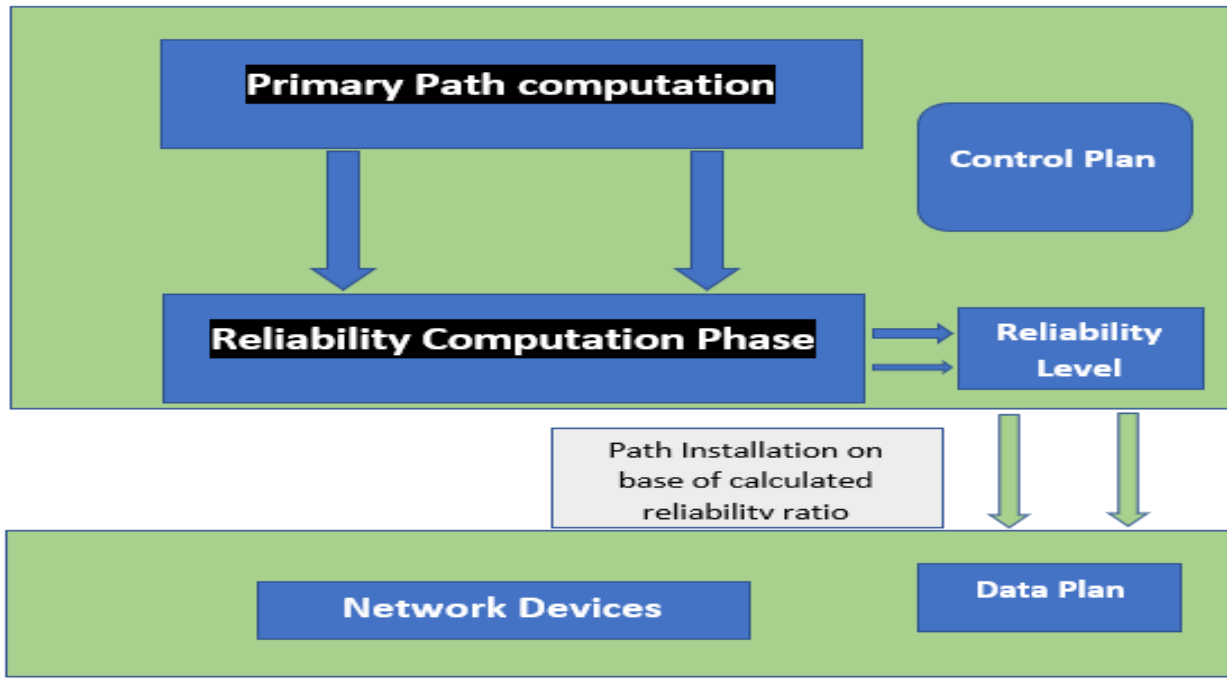

*Figure 4.1 Abstract View of Reliability Aware Flow installation*

### 4.1.1. Bootstrapping Process

When the network is switched-on, Hello, Feature-Request, and Feature-Reply [45] messages of OpenFlow protocol are exchanged between a switch and the controller in order to get the network view at the controller. In the proposed work, through Hello messages, the controller periodically generates a Feature-Request message to get the features of the switch. After receiving Feature-Request message, the switch replies to the controller by sending a Feature-Reply message. Along with other information, switch feature reply message also contains the reliability level of a link.

### 4.1.2. Path computation

As we mentioned above that we assume a reactive flow installation mode. In this mode, when the network is configured and starts the running, the flow tables at the switches are empty. When the data packet of a flow arrives at the switch, the switch looks for the matching entry in its forwarding table. If the matching is found, then the switch forwards the data packet according to the corresponding action of the flow table entry. Otherwise, the switch asks controller to compute the action for the flow. After receiving the request from the switch, the controller checks the access control list (ACL) whether the flow is allowed or denied. If the flow is denied, then the controller installs the drop action at the switches. Otherwise, the controller computes the of primary path using either of the following approaches

### 4.1.1. RAF

In RAF, we compute the most reliable path as primary path and follows either of the following cases for computing and installing other alternative paths RAF.

**Case I:**
If the reliability of primary path is more than 90%, RAF does not compute and install any other alternative path.

**Case II:**
If the reliability value of primary path is between 80% and 90 %, the controller installs two alternative paths.

**Case III:**
If the reliability value of primary path is between 70% and 80 %, the controller installs three alternative paths.

**Case IV:**
If the reliability value of primary path is between 60% and 70 %, the controller installs four alternative paths.

**Case V:**

If the reliability value of primary path is between 50% and 60 %, the controller installs five alternative paths.

**Case VI:**

If the reliability value of primary path is between 0 % and 50 %, the controller installs all available alternative path.

### 4.1.2. Distance based RAF

Distance based RAF considers computes the primary path based on the joint value of higher reliability and shorter path length. After this the controller computes and installs multiple path based on the reliability value of the primary path using Case I to Case VI in Section 4.1.1.

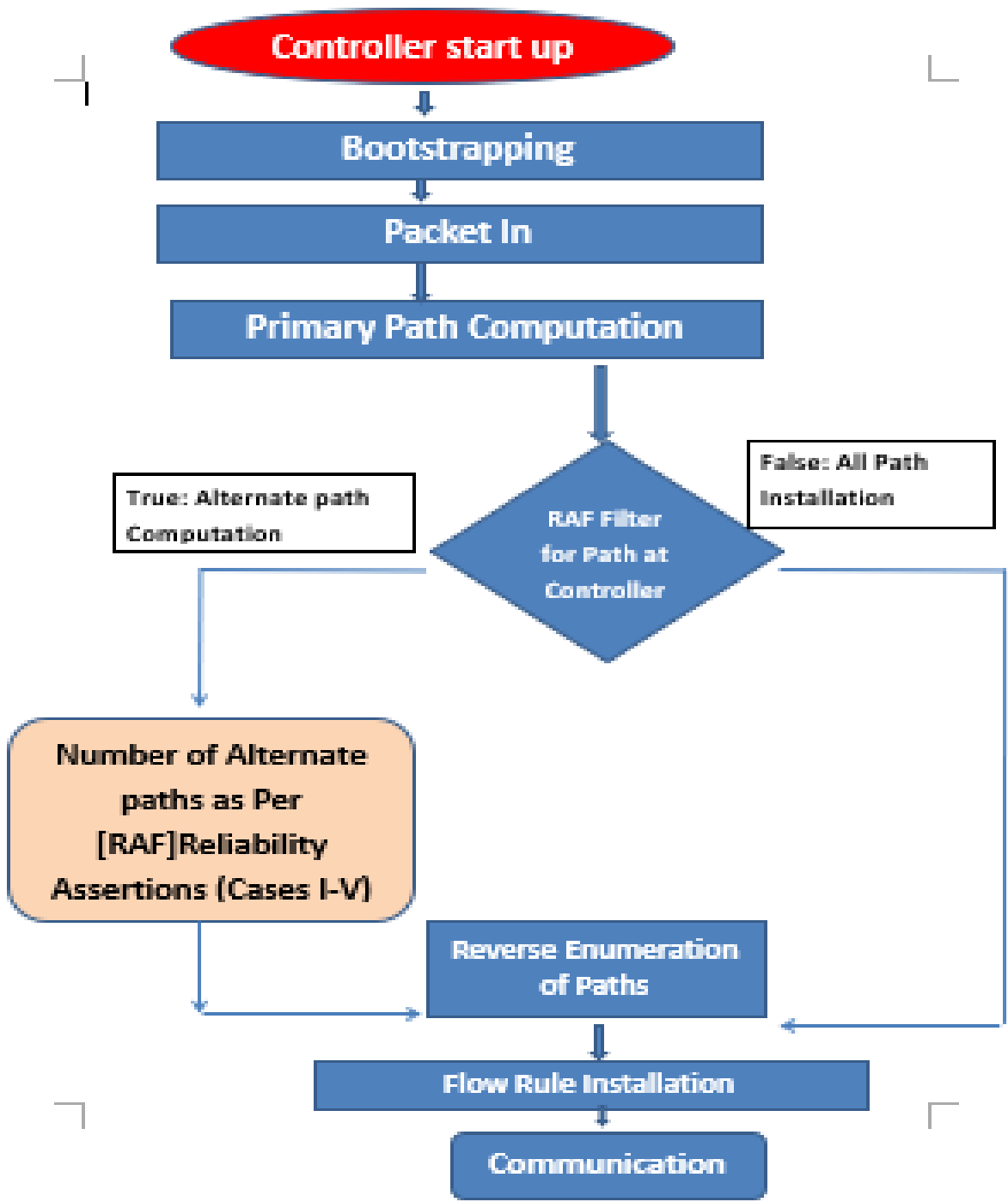

*Figure 4.2-Flow Diagram Of RAF Functionality in Reactive SDN*

Flow diagram 4.3 present the RAF methodology of reliability aware flow rule installation. RAF works in reactive flow installation mode. Diagram presents that a network started form controller and followed by a bootstrapping process in order to get the all network view of switches connectivity when controller find a primary reliable path ,If reliability level of primary path meets the assertion of installing only single path for controller rid to overhead and less flow rule entries as objective , controller shall enumerate the path reversely and install flow rule. Otherwise RAF residing at controller shall compute the number of alternate paths for recovery as per RAF percentage case assertions.

Paths are installed in data plane by installing the flow rules in the forwarding tables of the switches along the paths by using flow mod command of OpenFlow specifications. OFPFC_ADD is OpenFlow command used flow rule installation in flow table of switch. Match fields encapsulated in flow adding command are first compare for a corresponding flow rule entry in switch. Matching objects are match. dl_type(opcode of IPV4 , ARP etc. ) ,match.nw_proto (application layer protocol ) , match.nw_src (source IP address) , match.nw_dst(destination IP address). Controller also specify the priority and timeout values along with flow rule installation command. Traffic follow the high priority flow rules stored in switch memory in case of multiple flow rule for same source and destination When controller complete flow installation communication starts as per controller allowed decisions for respective protocols

### 4.1.2 Link Failure Process

Proposed method has functionally to make unavailability of path in case if link failure event occurs. Link failure is configured at switch egress ethernet interface   by OpenFlow command at controller. Controller adds a flow rule entry for binding the physical Ethernet interface to a logical port or simple setting the port status down. Failure procedure is activated when aliveness time of a link approaches up to zero. Controller gets information of failure using feature reply periodic message and makes the decision of alternative path installation.

# Chapter 5
# Simulation and Experimental Results

## 5.1 Simulation Tools

To evaluate the performance of proposed approaches, we have used the POX controller and Mininet simulator.

### 5.1.1 Mininet Simulator

With a need to fasten the research in Open flow and SDN Mininet emulator has been created that allows simulation for both small and large networks both without modification. Large scale topologies of size up to hundreds and thousands of nodes can be created using Mininet and can be tested easily that consists of simple tools for command line and API. Mininet [46] gives ease of use to the user that offers easy creation of SDN elements, customizing, sharing and testing SDN networks. Figure 5.1 shows the Mininet emulated network which have host and connectivity like a real network. The elements include switches, hosts, links and controllers. Further it provides the separate virtual environment for each host for executing various application

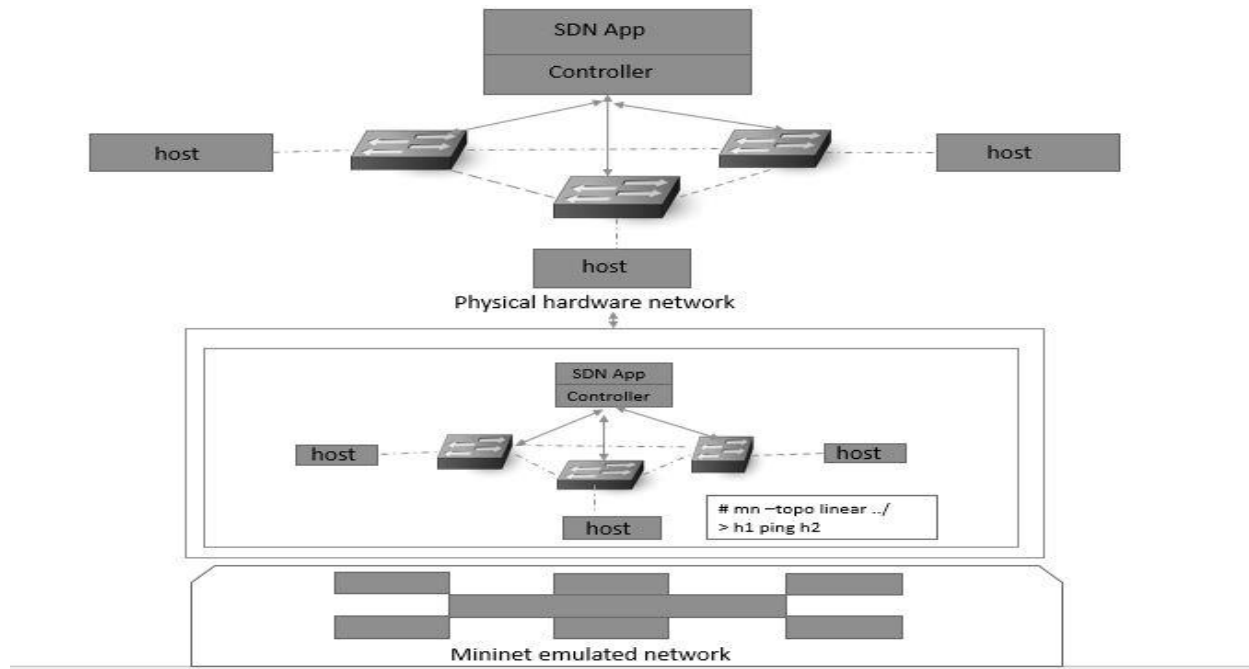

*Figure 5.1-Mininet Functionality and Actual SDN Network*

While simulating in Mininet, the controller can run on a real or simulated all the time. Machine that run the switches has connectivity via Ethernet interface to controller. The Mininet emulator draws connections between various types of controllers (like POX, NOX, Floodlight etc.) and switches, which allows the developers take part in creating and testing new resources that helps using Mininet in performing their desire simulation. Over 100 researchers has been using Mininet in institutions like Purdue, UMass, NASA, Stanford, Princeton, ICSI, Berkeley and several universities in Brazil Mininet Support the Multi controller connectivity and control of data rate, delay explicitly and other host links properties [47].

## 5.1.2 POX Controller

POX provides a framework for communicating with stateless switches using OpenFlow. . Developers can use POX to create an SDN controller via python language. It is a popular tool for academia researching in software defined networks and network management applications. POX can be immediately used as basic SDN controller by using the stock components that come with its versions [48].. Figure 5.2 explain event handling model of POX controller event handling. Developers may create a more complex SDN controller by creating new POX components. Pox supports OpenFlow switch specification 1.0 and 1.3 nearly Further Advantages and properties of POX controller are below.

- Open Source
- Interface able with Mininet
- Event-Driven to Support the discovery and dynamic view to topology
- Event publisher, subscriber fashion to support customer module development and event registration

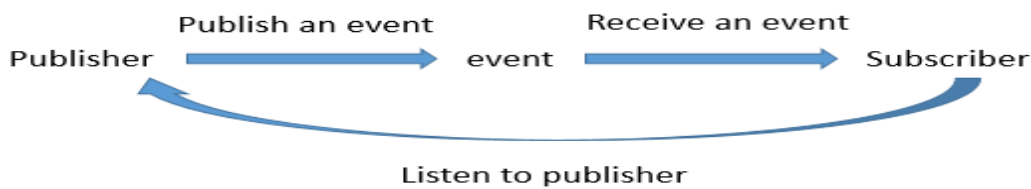

*Figure 5.2-Event Handling in POX controller*

### 5.1.3 OpenFlow API

OpenFlow is data plane devices management application programming interface that has used for testing our approach. OpenFlow was emerged in 2008 and continuously in updating situation. This API support user custom application and their decision to store in OpenFlow enabled switches via control plane commands. we have used its 1.0 version. 1.0 switch specification is supported at Pox controller.

### 5.1.4 Python

We have used python programming langue for management and data plane scripting , implementation of RAF as its compatibility with SDN controller and Mininet simulator. Related distributions, libraries , API  of python which are used for are Networkx (python graph composition) , Random(accessing iter able python object randomly )  , netaddr (Network Address Translation) , python socket API(TCP ,  UDP host communication interfaces  ) , Multithreading and Lock objects ( Daemon Threads  , Rlocks  , Semaphores ).

## 5.2 Experimental Results

We have 25 virtual end host and 9 OVS switches in our emulated network topology. Switches are interconnected. Host IP addressing is from IP subnet (10.0.0.1/16). A single host in this network topology generate 10000 UDP data packet with average of 62 bytes data per packet. Subsequent packet delay is controlling data rates of socket interfaces. Best results are collected via using 5 host generated data packets because of laptop limited resources.

Experiments are recursively performed using POX controller (2.2 eel version) , Mininet framework presents our work efficiency and concerned beneficial superiority comparative to existing approach of all path installation. We have all implementation scripts compiled in python language because of Pox and Mininet python 2.7 compatibility. Figure 5.3 present the Controller overhead for both of existing and proposed approach.

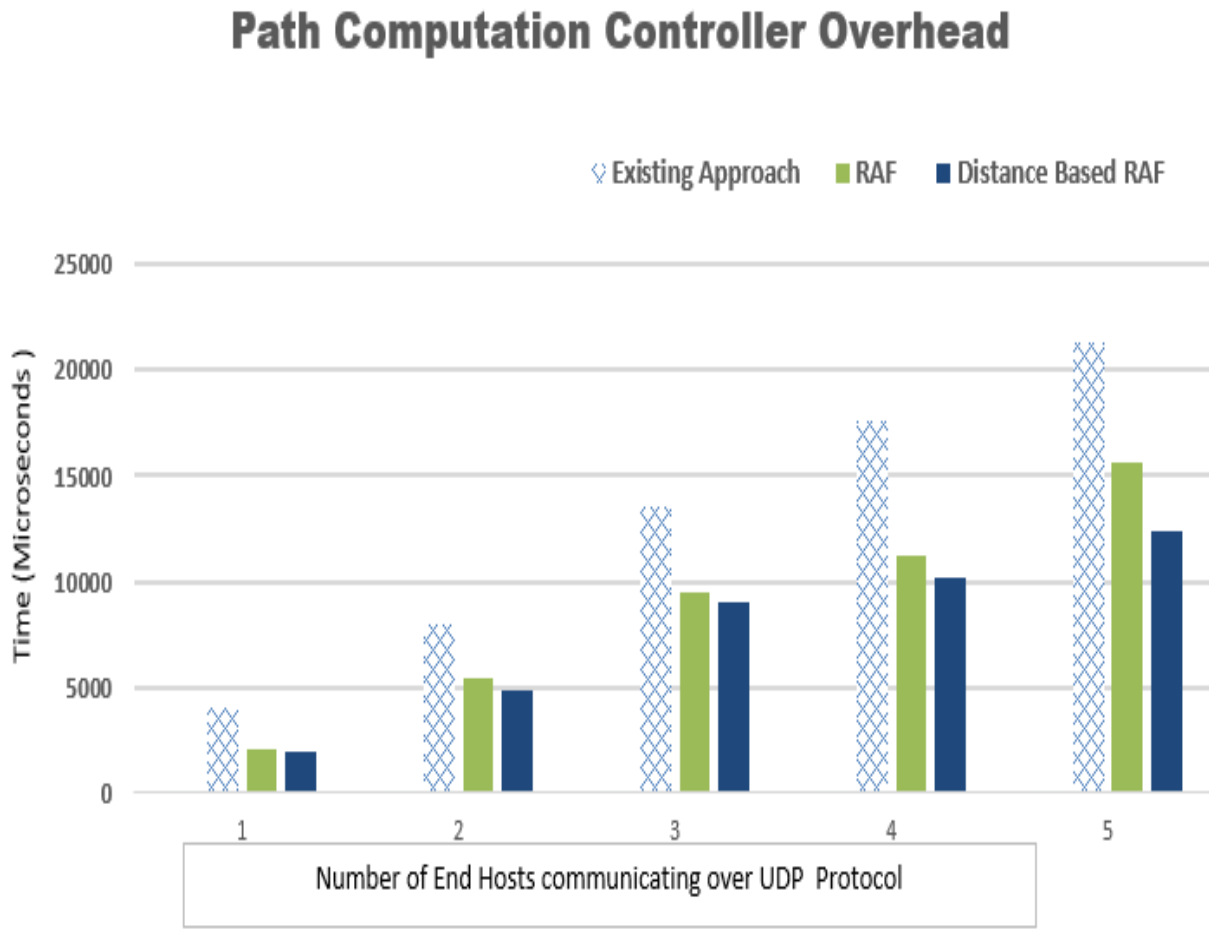

*Figure 5.3- Controller Overhead for Path Computation*

Average end-to-end delay of all path-based communication is higher than that of our RAF based communication. Actual reason behind this perspective of delay is controller high intervention of installing all path. During communication controller start multiple flow rule installation at egress port of switches that are intermediate devices of path from source to destination. Figure 5.4 elaborate our result of delay computation comparative efficiency because of streamlined communication and lesser controller involvement

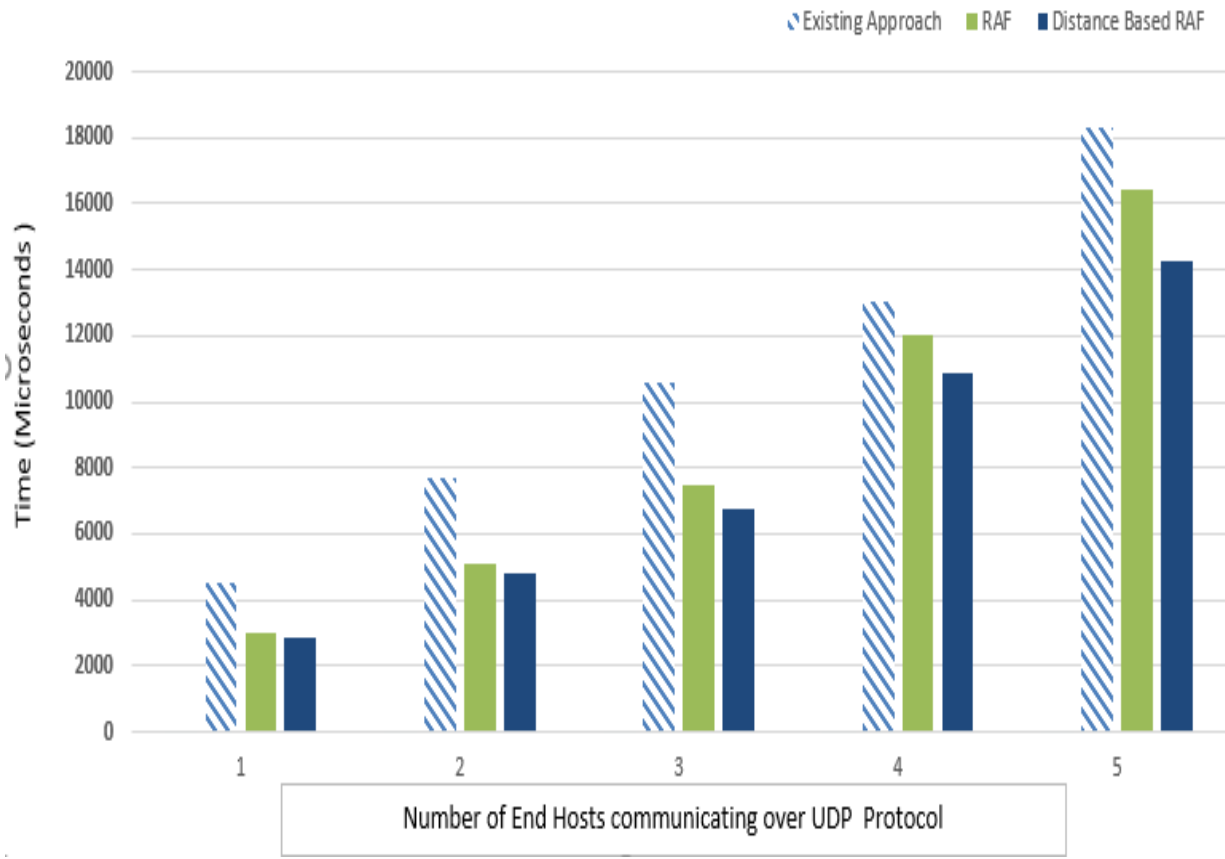

*Figure 5.4- Comparative average End-to-End Delay of existing and Reliable Path communication*

RAF minimize the memory frequent flow rule entries installation. When an OpenFlow enable device to receive frequent flow entries for installation its memory shows bottleneck. Figure 5.5 shows the number of flow rule installation in existing and RAF based approach.

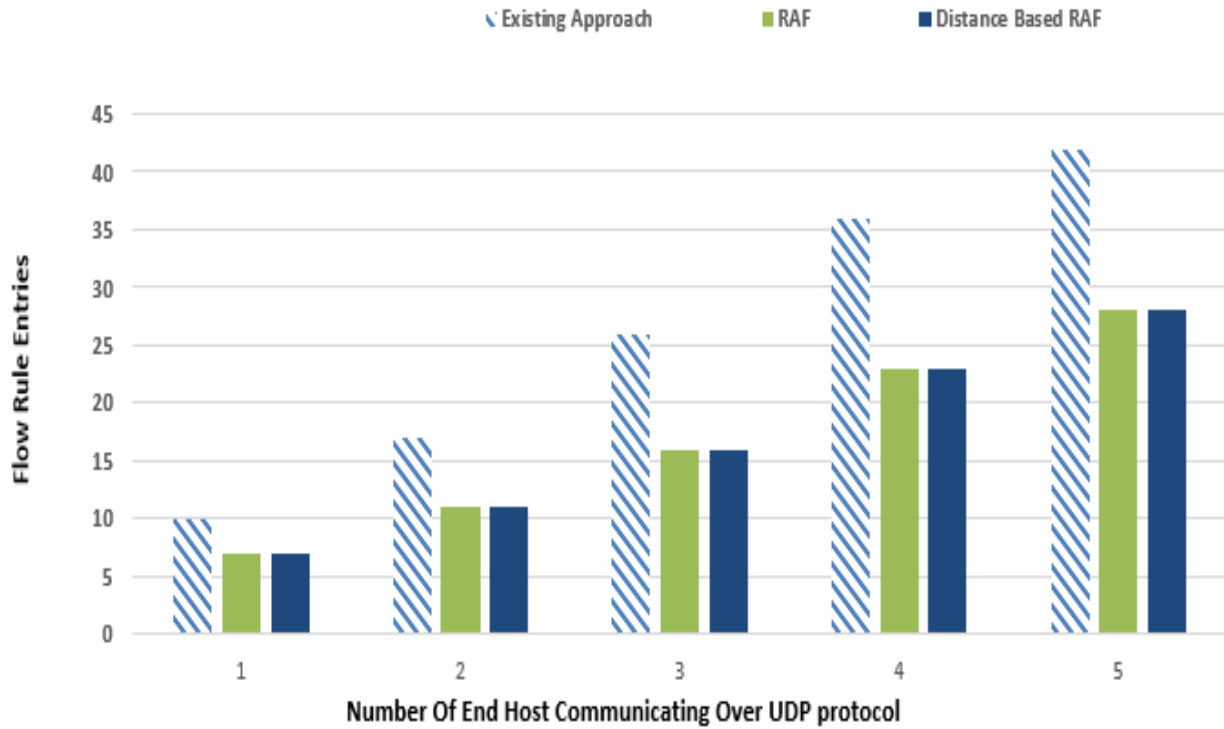

*Figure 5.5- OpenFlow enabled switch memory in existing and RAF approach*

As the Path length increase Controller has more and more control packet transmission rate. Figure 5.6 show the average transmission rate of OpenFlow messages in case all Path and for reliable paths. Existing approaches to much path installation load in control plane while RAF present an efficient memory usage trend. RAF and Distance based RAF shows equal number of flow rule because they have only difference of distance (path length ) awareness.

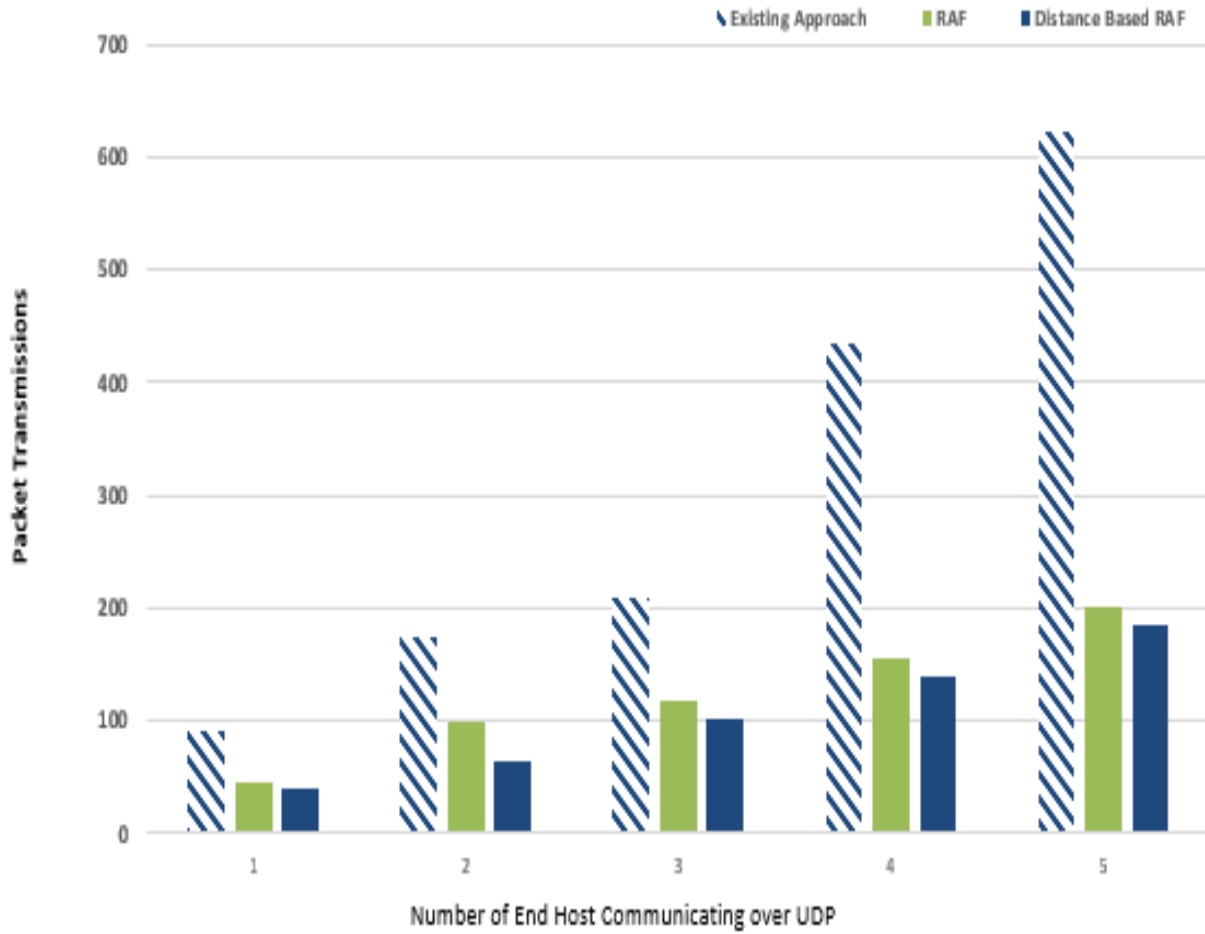

*Figure 5.6-Control Packet average Transmissions in existing approach and RAF based Flow installation*

# Chapter 6
# Conclusion

Software Defined Networking is an emerging computer network architecture by centralizing both the control and management plane. This centralization has many advantages like easy network control and management. Link failure occurs frequently in a network. To deal with the link failure, the existing approaches install all multiple paths in the network without considering the reliability value of the primary path. This causes more computation overhead at controller to compute all the paths, larger traffic overhead in the network to install all multiple paths at the switches in the network, more memory usage in the switches for install multiple paths. To address this problem, we have proposed a new approach, called RAF, which considers the reliability level of primary path and installs the number alternate paths according to the reliability value of the primary path. More specifically, if the primary path has higher reliability, then there is no need to install multiple paths. The simulation results confirmed that our proposed approach has better results by decreasing the computation load at controller, average end-to-end delay and the traffic overhead for installing multiple paths at data plane. As a future work, we would like to extend our proposed work by computing the link reliability using deep learning algorithms.

[30]. Gude, N., Koponen, T., Pettit, J., Pfaff, B., Casado, M., McKeown, N., & Shenker, S. (2008). NOX: towards an operating system for networks. ACM SIGCOMM Computer Communication Review, 38(3), 105-110.

[31]. Yu, Y., Xin, L., Shanzhi, C., & Yan, W. (2011, December). A framework of using OpenFlow to handle transient link failure. In Transportation, Mechanical, and Electrical Engineering (TMEE), 2011 International Conference o1n (pp. 2050-2053). IEEE.

[32]. Khurshid, A., Zhou, W., Caesar, M., & Godfrey, P. (2012, August). Veriflow: Verifying network-wide invariants in real time. In Proceedings of the first workshop on Hot topics in software defined networks (pp. 49-54). ACM.

[33]. Cascone, C., Sanvito, D., Pollini, L., Capone, A., & Sansò, B. (2017). Fast failure detection and recovery in SDN with stateful data plane. International Journal of Network Management, 27(2), e1957.

[34]. Stephens, B., Cox, A. L., & Rixner, S. (2016, March). Scalable multi-failure fast failover via forwarding table compression. In Proceedings of the Symposium on SDN Research (p. 9). ACM.

[35]. Guo, C., Wu, H., Tan, K., Shi, L., Zhang, Y., & Lu, S. (2008, August). Dcell: a scalable and fault-tolerant network structure for data centers. In ACM SIGCOMM Computer Communication Review (Vol. 38, No. 4, pp. 75-86). ACM.

[36].https://3vf60mmveq1g8vzn48q2o71a-wpengine.netdna-ssl.com/wp-content/uploads/2014/10/openflow-switch-v1.5.1.pdf

[37]. Cheng, Z., Zhang, X., Li, Y., Yu, S., Lin, R., & He, L. (2017). Congestion-aware local reroute for fast failure recovery in software-defined networks. IEEE/OSA Journal of Optical Communications and Networking, 9(11), 934-944.

[38]. Basuki, A. I., & Kuipers, F. (2018, August). Localizing link failures in legacy and SDN networks. In 2018 10th International Workshop on Resilient Networks Design and Modeling (RNDM) (pp. 1-6). IEEE.]

[39] Li, Q., Liu, Y., Zhu, Z., Li, H., & Jiang, Y. (2019). BOND: Flexible failure recovery in software defined networks. Computer Networks, 149, 1-12.]
56